\documentclass[nofootinbib,prl,
floatfix,preprintnumbers,
amsmath,amssymb,twocolumn
]{revtex4}
\usepackage{epsfig}
\usepackage[colorlinks,citecolor=blue]{hyperref}
\usepackage{amsmath}

\def\be{\begin{equation}}
\def\ee{\end{equation}}
\def\bea{\begin{eqnarray}}
\def\eea{\end{eqnarray}}
\begin{document}
\title{A universal correction to higher spin entanglement entropy}
\author{Shouvik Datta,$^a$ Justin R. David,$^{a}$ Michael Ferlaino$^{b}$ and S. Prem Kumar$^{b}$
}
\affiliation{
$^a$Centre for High Energy Physics, Indian Institute of Science, C.V. Raman Avenue, Bangalore 560012, India,\\
$^{b}$Department of Physics, Swansea University, Singleton Park, Swansea, 
SA2 8PP, U.K.}
\begin{abstract}
We consider conformal field theories in 1+1 dimensions with ${\cal W}$-algebra symmetries, deformed by a chemical potential $\mu$ for the spin-three current. We show that the order $\mu^2$ correction to the R\'enyi and entanglement entropies of a single interval in the deformed theory, on the infinite spatial line and at finite temperature, is universal. The correction is completely determined by the operator product expansion of two spin-three currents, and by the expectation values of the stress tensor, its descendants and its composites, evaluated on the $n$-sheeted Riemann surface branched along the interval. This explains the recently found agreement of the order $\mu^2$ correction across distinct free field CFTs and higher spin black hole solutions holographically dual to CFTs with ${\cal W}$-symmetry. 
\end{abstract}
\maketitle

{\it Introduction.$-$} Conformal field theories (CFTs) in two dimensions admit non-trivial extensions of the infinite-dimensional Virasoro conformal symmetry to ${\cal W}$-algebra symmetries, generated by conserved currents with spins $s\geq 2$.  The ${\cal W}$-symmetries not only play a powerful role in the analysis and classification of 2d CFTs, but also appear in appropriate scaling limits of statistical systems e.g. the three-state Potts model and more general $Z_N$ lattice models \cite{fz}. It has emerged recently that CFTs with ${\cal W}$-symmetry are holographically dual to higher spin theories of gravity on AdS$_3$ \cite{asymp, gg}. The proposal of Gaberdiel and Gopakumar \cite{gg} specifically relates ${\cal W}_N$ minimal model CFTs in a 't Hooft large-$N$ limit to Vasiliev's higher spin theory of gravity on AdS$_3$ \cite{Prokushkin:1998bq}. Such dualities between higher spin theories and conformal field theories present us a fascinating framework within which to  better understand gauge-gravity dualities \cite{maldacena} at the `nuts and bolts' level. As a bonus we  also expect them to shed some light on the nature of gravity in string theory when  spacetime curvatures become large and consequently the tower of higher spin string excitations becomes light. 

Of particular interest is the thermodynamics of ${\cal W}$-algebra CFTs since this is directly connected, via holographic duality, to the thermodynamics of black hole solutions in higher spin gravity. Black hole solutions in AdS$_3$, carrying higher spin charge, have been constructed \cite{gutkraus} and their thermodynamics shown to match that of CFTs with chemical potential for higher spin charge \cite{fromCFT}. A deeper understanding of this holographic duality requires more refined probes of the CFT and correspondingly, of its gravity dual. The entanglement entropy of the CFT with chemical potential for higher spin charge is one such observable which goes beyond thermodynamics, and probes further aspects of the black hole solutions in higher spin gravity. Below, we will argue that the first non-trivial correction  to the R\'enyi entanglement entropy of a single interval, in a CFT with  ${\cal W}$-symmetry with a chemical potential for spin-three charge, is actually universal.

For a generic CFT in two dimensions with ${\cal W}$-symmetry (working in Euclidean signature) we will only assume that the theory is endowed with a stress tensor $T(z)$ and a spin-three primary current $W(z)$  (along with anti-holomorphic counterparts), and that the CFT is deformed as
\be
{\cal L}_{\rm CFT}\to {\cal L}_{\rm CFT} - \left(\mu\,W(z) \,-\bar \mu \,\overline W(\bar z)\right)\,.
\ee
We take $\mu=-\bar\mu$ motivated in part by the spin-three black hole solution of \cite{gutkraus} where the two sectors are treated symmetrically. However, the results in this paper do not depend on this choice and can be carried over straightforwardly to the situation with $\mu\neq-\bar \mu$.
The coupling $\mu$ can be viewed as a chemical potential for the spin-three charge. The holomorphic current $ W$ has scaling dimension three, so the deformation is irrelevant at first sight. Nevertheless, given that the deformation is by a $(3,0)$ operator, the resulting perturbation theory in $\mu$ is non-standard, and appears well-defined with no new divergences \cite{fromCFT, paper1}. We are interested in calculating the entropy of entanglement $S_{\rm EE}(\Delta)$, of a single spatial interval of length $\Delta$, for the deformed CFT on a line and at any temperature $\beta^{-1}$. The well known procedure for evaluating this is to perform the replica trick to compute the partition function of the CFT on an $n$-sheeted Riemann surface ${\cal R}_n$, branched along the interval, first yielding the so-called R\'enyi entropies(RE) \cite{cardycalab}
\be
S^{(n)}\,=\, \frac{1}{1-n}\ln\left[\frac{Z^{(n)}}{Z^n}\right]\,,\qquad S_{\rm EE}\,=\,\lim_{n\to 1}S^{(n)}\,.\label{RE}
\ee
Here $Z^{(n)}$ and $Z$ are the CFT partition functions on the branched Riemann surface and the complex plane respectively. The R\'enyi and entanglement entropies are formally related to the reduced density matrix $\rho_\Delta$ for the interval $\Delta$ as,
$S^{(n)}=\ln{\rm Tr}\rho_\Delta^n/(1-n)$ and $S_{\rm EE}\,=\,-{\rm Tr}{\rm\rho_\Delta }\ln\rho_\Delta$.
As pointed out in \cite{cardycalab}, the calculation of $Z^{(n)}$ may be achieved by evaluating the partition function in the presence of {\em twist} and {\em anti-twist} fields ${\cal T}_n$ and $\overline{\cal T}_n$ inserted at the end-points of the interval, so that in a perturbative expansion in $\mu$, we have
\bea
&& S^{(n)}\,=\,\frac{1}{1-n}\,\ln Z^{-n}\left(Z_{\rm CFT}^{(n)}\,+\,\right.\\\nonumber 
&&\left.\frac{\mu^2}{2}\int d^2 z_1 d^2 z_2 \langle\overline{\cal T}_n(y_2)W(z_1)W(z_2){\cal T}_n(y_1)\rangle+ {\rm h.c.}\ldots\right)
\eea
with $y_1, y_2$ real and $|y_2-y_1|=\Delta$.
If $W$ is primary with vanishing one-point function on the complex plane, 
under the uniformizing conformal map from the plane to  ${\cal R}_n$ (discussed below) we will continue to have $\langle W\rangle_{{\cal R}_n} = 0$. The mixed two-point function $\langle W(z)\overline W(\bar z) \rangle_{{\cal R}_n}$ would  also be vanishing, except for possible contact terms\footnote{We thank an anonymous referee for pointing this out.}. Such contact terms, which are products of delta-functions and their derivatives, will be dealt with using an appropriate prescription for all the relevant integrals and will not contribute.
The first non-trivial correction to the R\'enyi entropies in the deformed theory is therefore determined by the four point correlator involving the twist-anti-twist pair and two $W$-currents. Importantly, the twist operators and their action can be made explicit only for free theories and consequently the correlators explicitly evaluated for these examples \cite{paper1}. We will now argue that the singularity structure of the correlator, relevant for RE evaluation, is determined in complete generality by the $WW$ operator product expansion (OPE).

{\it Four-point function.}$-$ By conformal invariance, the required correlator on ${\cal R}_n$ can be expressed in terms of an undetermined function of the conformal cross-ratio,
\bea
&&\langle \overline{\cal T }_n(y_2) W(z_1) W(z_2) {\cal T}_n(y_1)\rangle = -
\frac{5c/6\pi^2\,F(x)}{(z_{12})^{6}\,\Delta^{2 d_n}}\nonumber\\
&&\Delta\equiv|y_2-y_1|,\qquad x \equiv \frac{(z_1-y_2) (z_2- y_1)}{(z_1-y_1)(z_2-y_2)}\,, 
\label{4pt}
\eea
where $z_{12}\equiv(z_1-z_2)$, $d_n\,=\, c\, (n-1/n)/12$ is the conformal dimension of the twist operators \cite{cardycalab} and $c$ the central charge of the CFT. The unknown function $F(x)$ satisfies the following properties:
\begin{itemize}
\item{$F(x)$ is a {\em holomorphic} function of $x$.}
\item{ Symmetry under $z_1\leftrightarrow z_2$ implies $F(x)= F(1/x)$. Symmetry under exchange of the endpoints of the interval $y_1\leftrightarrow y_2$ is then automatic.}
\item{The expansion of $F(x)$ near $x=1$, as $z_1\to z_2$, is determined by the $WW$ OPE normalized so that $F(1)=1$.}
\item{When either of the two $W$-currents approach a twist field, we have $F(x)\sim x^M$ as $x\to \infty$ and $F(x)\sim x^{-M}$ as $x\to 0$.}
\end{itemize}
 The value of $M$ is restricted by the OPE of the spin-three current with the branch point twist fields,
\be
W(z){\cal T}_n(y)\sim \frac{1}{(z-y)^M}\,{\widetilde {\cal T}}_n(y)\,+\ldots,
\ee
where $\widetilde{\cal T}_n$ is an excited twist operator obtained by the action of an untwisted field on the twisted sector vacuum. A possible term  $\sim {\cal T}_n/(z-y)^3$ in the OPE is forbidden as it would imply a one-point function for the primary field $W$ on ${\cal R}_n$.
By dimensional analysis, since the excited twist operator has scaling dimension larger than that of ${\cal T}_n$, we must have that $M < 3$. In fact, the explicit calculation of the OPE between ${W}$ and ${\cal T}_n$ for the free boson and free fermion CFTs \cite{paper1} shows that in both cases $M=2$. This is because $W(z)$ continues to be integer moded (in free field realizations this is a consequence of  currents being bilinears in the elementary fields), and the ground state has $\langle W\rangle_{{\cal R}_n} =0$. This can be understood from the general idea underlying the replica approach, wherein the Lagrangian density (including deformations) is represented as a sum over $n$ independent copies $\delta {\cal L}=-\mu W\,=\,-\mu \sum_{i=1}^n W_{(i)}$. Since the action of the twist fields on the replicas is by cyclic permutation of the copies, it is a symmetry of the Lagrangian and of the deforming operator $W$, and hence the OPE of the twist fields with $W$ does not contain branch cuts \cite{cardycalab}.

Finally, the absence of branch points at $x=0$ and $x=\infty$ automatically renders $x=1$ a regular point (since a branch cut originating at $x=1$ needs to end at another distinguished point). 
All the requirements above specify the form of $F$ uniquely as
\bea
&&{F}\,=\,1+{f}_1\,\eta + f_2 \eta^2\,,\label{etaexp}\,\\\nonumber
&&\eta\equiv x+\frac{1}{x}-2=\frac{(z_1-z_2)^2(y_1-y_2)^2}{(z_1-y_1)(z_1-y_2)(z_2-y_1)(z_2-y_2)}\,.\nonumber
\eea
While the leading singularity of the four-point correlator, as $z_1 \to z_2$, is controlled by the conformal dimension of $W$ and the central term in the $WW$ OPE below, the subleading singular terms are dictated by the two coefficients $f_1$ and $f_2$.

It is instructive to trace the corresponding  arguments for the stress tensor $T$ whose two-point function in the presence of branch-point twist fields can also be determined universally using these methods. The main difference is that the stress tensor acquires a one-point function due to the OPE,
\bea
T(z){\cal T}_n(y)\sim\frac{d_n}{2}\frac{{\cal T}_n(y)}{(z-y)^2}+\frac{{\cal T}_n'(y)}{(z-y)}\,.\label{twistt}
\eea
We then find that the corresponding correlator is completely constrained to be of the form,
\bea
\langle \overline{\cal T }_n(y_2) T(z_1) T(z_2) {\cal T}_n(y_1)\rangle=\frac{c\left(1+ g_1\eta + g_2 \eta^2\right)}{2\,{z_{12}^4}\,\Delta^{2d_n}}\,\label{stress4pt}
\eea
where $g_1$ and $g_2$ will be pinned down below.

{\it ${\cal W}$-algebra OPEs.}$-$ The relevant OPEs of the ${\cal W}_\infty[\lambda]$ algebra (including  the finite ${\cal W}_N$-algebras for $\lambda=-N$) are \cite{gabjin} 
\bea
&&T(z_1) T(z_2) \sim \frac{c}{2\,z_{12}^4} + \frac{2T(z_2)}{z_{12}^2} + \frac{\partial T(z_2)}{z_{12}}\\\nonumber
&&T(z_1) W(z_2)  \sim \frac{3 W(z_2)}{z_{12}^2} +\frac{\partial W(z_2)}{z_{12}}\,,\\\nonumber
&&\tfrac{1}{N_3}W(z_1) W(z_2)\sim\frac{5 c/6}{z_{12}^6}
+\frac{5  T(z_2)}{z_{12}^4}+\frac{5  T'(z_2)/2}{ z_{12}^3} \nonumber \\\nonumber
&&+\frac{1}{z_{12}^2}\left({4 U(z_2)+\frac{16 }{c+22/5}\Lambda^{(4)}(z_2)+\tfrac{3}{4}  T''(z_2)}\right)
\\\nonumber
&&+\frac{1}{z_{12}}\left({2\partial U(z_2)+\frac{8}{c+22/5} \partial\Lambda^{(4)}(z_2)+\tfrac{1}{6} T'''(z_2)}\right)\,. \label{ww-ope}
\eea
We have taken the spin-four current $U(z)$ to be a primary field. The composite operator $\Lambda^{(4)}$ and the normalization constant $N_3$ are defined as
\be
\Lambda^{(4)}=:TT: -\tfrac{3}{10}\partial^2T\,,\qquad
N_3=-\pi^{-2}\,.
\ee
Therefore the coefficients of singular terms (as $z_1\to z_2$) in the correlator $\langle \overline {\cal T}_n WW {\cal T}_n \rangle$ are determined by the one-point functions $\langle T\rangle_{{\cal R}_n}$, $\langle\partial^p T\rangle_{{\cal R}_n}$ for $(p=1,2,3)$ and $\langle:TT:\rangle_{{\cal R}_n}$. The one-point function of $U$ must vanish on ${\cal R}_n$ if we take it to transform as a primary. Note that whilst {\em a priori} there are {\em four} independent one-point functions in  the OPE, there are only {\em two} parameters $f_{1,2}$ controlling the singularity structure of the four-point correlator $\langle\overline{\cal T}_nWW{\cal T}_n\rangle$. The system is, however, not overdetermined and we obtain a unique, consistent solution for the pair $(f_1, f_2)$.

{\it The uniformization map.}$-$ All requisite one-point functions for the stress tensor, its descendants and associated composites, can be obtained by application of the uniformizing conformal transformation which maps the Riemann surface ${\cal R}_n$ to the complex $w$-plane ${\cal C}_w$:
\bea
w= \left( \frac{z-y_2}{z-y_1}  \right)^{1/n}\,. \label{zwmap}
\eea
When the CFT on the spatial line is at a temperature $\beta^{-1}$, the relevant map  from the branched $n$-sheeted cylinder   
to the $w$-plane is achieved by setting 
\be
z=e^{2\pi u/\beta}\,,\qquad0\leq{\rm Im}(u)<\beta\,,
\label{cylmap}
\ee
in eq.\eqref{zwmap}, where $u$ is the co-ordinate on the cylinder. For the sake of clarity we will quote all intermediate steps for the $n$-sheeted cover ${\cal R}_n$ of the complex plane and in the end write the result for the cylinder using the map \eqref{cylmap}.
Under a conformal transformation the stress tensor transforms as
\be
T(z)=w'(z)^2\,T(w) +\tfrac{c}{12}\{w,z\}\,,\label{ttrans}
\ee 
where $\{w,z\}=(w'''\,w'-\tfrac{3}{2}{w''}^2)/{w'}^2$ is the Schwarzian for the map.
The stress tensor has vanishing one-point function on the $w$-plane, $\langle T(w)\rangle_{{\cal C}_w} =0$, by translation and rotational invariance. Hence the one-point function on ${\cal R}_n$ is given by the Schwarzian,
\be
\langle T(z)\rangle_{{\cal R}_n}\,=\,\tfrac{c}{12}\{w,z\}\,=
\frac{c\,(n^2-1)\,\Delta^2}{24\,n\,(z-y_1)^2(z-y_2)^2}\,.
\ee
We note that whilst evaluating all one-point functions, one must include a factor of $n$ to account for contributions from all $n$ copies or Riemann sheets of ${\cal R}_n$. 
The one-point functions $\langle\partial^p T(z)\rangle_{{\cal R}_n}$ automatically follow by taking derivatives of this result. The new ingredient required is the one-point function of $:T(z)T(z):$ or equivalently, that of the composite operator $\Lambda^{(4)}$ on ${\cal R}_n$. This can be obtained using the point-split definition of the normal ordered product 
\bea
&&:T(z)T(z):=\,\lim_{\epsilon\to 0}\left(T(z+\tfrac{\epsilon}{2})T(z-\tfrac{\epsilon}{2})\right.\\\nonumber
&&\left.\qquad\qquad\qquad-\frac{c}{2\epsilon^4} -\frac{2T(z-{\epsilon}/{2})}{\epsilon^2}-
\frac{\partial T(z-{\epsilon}/{2})}{\epsilon}\right)\,
\eea
and then applying the transformation law \eqref{ttrans} for the stress tensor under the uniformization map. We find
\be
\langle\Lambda^{(4)}\rangle_{{\cal R}_n}= \frac{c(5c+22)}{2880}\frac{(n^2-1)^2\,\Delta^4}
{n^3\,(z-y_1)^4(z-y_2)^4}\,.
\ee
Expanding the cross-ratios in \eqref{etaexp} in a Laurent series in $(z_1-z_2)$, we obtain perfect agreement with all the 
singular terms in the OPE on ${\cal R}_n$ provided
\be
f_1=\frac{n^2-1}{4n}\,,\qquad f_2=\frac{(n^2-1)^2}{120 n^3}-\frac{(n^2-1)}{40n^3}\,.
\ee
This coincides with the result of explicit calculations in \cite{paper1} for the free boson theory with ${\cal W}_{\infty}[1]$ symmetry. 
Therefore, the four-point function \eqref{4pt} is fixed universally for any CFT with ${\cal W}_\infty[\lambda]$ symmetry, including the special cases with a finite 
${\cal W}_N$-algebra (for integer $N> 2$). In all cases, the primary spin-four current with vanishing one-point function does not contribute, and the correlator is determined by the ${\cal W}_3$ subalgebra. 

{\it ${\cal W}_{1+\infty}$ and free fermions}$-$ The discussion above pertains to theories with ${\cal W}_{\infty}[\lambda]$ symmetry. One of the simplest theories with a higher spin symmetry $-$ the free fermion CFT, has a slightly different symmetry algebra namely ${\cal W}_{1+\infty}$, generated by a spin-1 current $J(z)$, the stress tensor and higher spin currents. The spin-three and spin-four currents in the free fermion representation are {\em not} primaries and the $WW$ OPE has the form 
\bea
&&\tfrac{1}{N_3}W(z_1) W(z_2)\sim\frac{5 c/6}{z_{12}^6}+\frac{5  T(z_2)}{z_{12}^4}+\frac{5  T'(z_2)/2}{ z_{12}^3} 
\\\nonumber
&&\qquad+\frac{5 U(z_2)+\frac{3}{4}  T''(z_2)}{z_{12}^2} 
+\frac{\frac{5}{2}\partial U(z_2)+\frac{1}{6} T'''(z_2)}{z_{12}}\,. \label{ww-ope}
\eea
$W(z)$ transforms as a primary only in the $J=0$ sector, and therefore has vanishing  one-point function on ${\cal R}_n$ in this sector. On the other hand, the spin-four current $U(z)$ whose explicit form can be found in \cite{gabjin}, obtains a one-point function even in the $J=0$ sector and we find,
\be
\langle U\rangle_{{\cal R}_n}=\frac{7c}{960}\frac{(n^2-1)^2\Delta^4}{n^3(z-y_1)^4(z-y_2)^4}\,.
\ee
Comparison of the short distance behaviour from the OPE and the cross-ratio expansion then yields
\be
f_1=\frac{n^2-1}{4 n}\,,\qquad f_2=\frac{3(n^2-1)^2}{160n^3}-\frac{(n^2-1)}{40n^3}\,.\label{ff}
\ee
This reproduces the result of explicit computations in the free fermion CFT \cite{paper1}. Interestingly, only the coefficient $f_2$  in \eqref{ff} differs from that for 
${\cal W}_\infty[\lambda]$ theories and furthermore the difference involves a term that does not contribute in the $n\to 1$ limit relevant for entanglement entropy \eqref{RE}.

{\it Stress tensor correlator.}$-$ For completeness and for comparison we also present the result for the  correlator $\langle{\overline{\cal T}_n TT{\cal T}_n}\rangle$ which is determined by the coefficients $g_1$ and $g_2$ in \eqref{stress4pt}. Whilst $g_1$ is fixed by the $TT$ OPE and $\langle T\rangle_{{\cal R}_n}, \langle T'\rangle_{{\cal R}_n} $, the second coefficient $g_2$ follows from the $T{\cal T}_n$ OPE \eqref{twistt}:
\bea
&& g_1=\frac{n^2-1}{6n}\,,\qquad\qquad\qquad g_2=\frac{c(n^2-1)^2}{288 n^3}\\\nonumber
&&{\Delta^{2d_n}}\langle\overline{\cal T}_n(y_2) T(z_1)T(z_2){\cal T}_n(y_1)\rangle\,=\,\\\nonumber
&&\qquad\quad\left(\frac{c}{2z_{12}^4}+c\,g_1\frac{\eta}{2z_{12}^4} + \frac{1}{n}\langle T(z_1)\rangle_{{\cal R}_n}\langle T(z_2)\rangle_{{\cal R}_n}\right)\,.
\eea
We have obtained the same result by applying conformal Ward identities to the three-point function i.e. by acting with infinitesimal conformal transformations on $\langle\overline{\cal T}_n(y_2)T(z_2){\cal T}_n(y_1)\rangle$.

{\it R\'enyi entropies.$-$} We now outline the calculation of the order 
$\mu^2$ correction to the R\'enyi entropy for the CFT on the line, at finite temperature. To begin with, the first term $(\sim \eta^0)$ in the four-point correlator \eqref{4pt} yields the correction to the thermal free energy density $\sim\int d^2u_1\sinh^{-6}(\pi u_{12})/\beta)$ at order $\mu^2$, where $u_{12}\equiv u_1-u_2$. This term is cancelled  in the ratio $Z^{(n)}/Z^n$. In particular, the thermal entropy density at order $\mu^2$ is
\be
L^{-1}S_{\rm thermal}=\frac{c}{3}\pi \beta^{-1}\,+\,\frac{32\pi^3}{9}\mu^2\beta^{-3}+\ldots\label{thermal}\,,
\ee
where $L$ signifies the size of the system.
On the finite temperature cylinder, the conformal cross ratio becomes, using the conformal transformation \eqref{cylmap}
\bea
&&\eta\to\eta_{\beta}= \frac{\sinh^2\left(\tfrac{\pi}{\beta}(u_1-u_2)\right) \,\sinh^2\left(\tfrac{\pi}{\beta}\Delta\right)}{\sinh\left(\tfrac{\pi}{\beta}(u_1-y_1)\right)\sinh\left(\tfrac{\pi}{\beta}(u_1-y_2)\right)}
\times\nonumber\\
&&\times\frac{1}{\sinh\left(\tfrac{\pi}{\beta}(u_2-y_1)\right)\sinh\left(\tfrac{\pi}{\beta}(u_2-y_2)\right)}\,,
\eea
where $y_{1,2}$ continue to denote the endpoints of the entangling interval on the real line with $|y_1-y_2|=\Delta$. Thus, after accounting for both holomorphic and anti-holomorphic sectors, the finite temperature R\'enyi entropies at order $\mu^2$ are given by
\bea
&&S^{(n)}=\frac{c(1+n)}{6n}\ln\left|\sinh\left(\tfrac{\pi}{\beta}\Delta\right)\right|
+ S^{(n)}_2+\ldots\label{renyi}
\\\nonumber
&&
S^{(n)}_2=\frac{5\pi^4c\mu^2}{6\beta^6(n-1)}\int d^2u_1\int d^2 u_2\frac{(f_1\, \eta_\beta+f_2\eta_\beta^2)}{ \sinh^{6}\left(\frac{\pi}{\beta}u_{12}\right)}\,.
\eea
The result of the integrals over the cylinder in \eqref{renyi} has been presented in detail in \cite{paper1}. The key property that facilitates their evaluation is that the integrands are holomorphic and bounded at infinity on the cylinder. 
For such functions, the integral is governed by the coefficients of the simple and double pole singularities of the integrand. 

A particularly subtle feature of the integrals encountered above is the presence of contact term singularities which arise when $u_1$ and $u_2$ collide with each other or with any of the branch point twist fields at $y_1$ and $y_2$. We must provide a prescription for dealing with such potential singularities. We motivate the  prescription we adopt (as in \cite{paper1}) by explaining how it applies to the simpler case of the order $\mu^2$ correction to the thermal partition function. The relevant integral and our prescription for evaluating it is as follows ($u_{1,2}\equiv
\sigma_{1,2}+i\tau_{1,2}$)
\begin{eqnarray}
&&\int d^2 u_1\int d^2 u_2\langle W(u_1)W(u_2)\rangle\,\equiv\\\nonumber
&&\lim_{\epsilon\to 0}
\int_0^\beta d\tau_2\int d\sigma_2\int d\sigma_1\left(\int_{0}^{\tau_2-\epsilon}d\tau_1
\langle W(u_1) W(u_2)\rangle\right.
\nonumber\\\nonumber
&&\left.\hspace{1.3in} + \int_{\tau_2+\epsilon}^\beta d\tau_1
\langle W(u_1) W(u_2)\rangle\right)
\end{eqnarray}
We now use the definition of the conserved charge as $Q_{\tau} = \int d\sigma\, W(\sigma+i\tau)$, where we have kept a $\tau$ label on the left hand side explicit, even though the charge is time independent. It is easy to check by explicit computation, that in all cases, the integrals are finite and independent of $\epsilon$ for any non-zero $\epsilon$. We emphasize that our prescription also requires performing the spatial integrals first.
 Formally taking $\epsilon$ to zero at the end, we can interpret the result as 
\begin{eqnarray}
&&\int d^2 u_1\int d^2 u_2\langle W(u_1)W(u_2)\rangle\,\equiv\\\nonumber\\\nonumber
&&\lim_{\epsilon\to 0} \int_0^\beta d\tau_2\left(\int_{0}^{\tau_2-\epsilon}d\tau_1
\langle Q_{\tau_1} Q_{\tau_2}\rangle +\int_{\tau_2+\epsilon}^\beta d\tau_1\langle Q_{\tau_1} Q_{\tau_2}\rangle\right)\nonumber\\\nonumber\\\nonumber
&&=\beta^2 \langle Q^2\rangle \,,
\end{eqnarray}
which precisely determines the expected correction to the canonical partition function. Other prescriptions for dealing with these contact terms (e.g.\cite{Douglas:1993wy, Dijkgraaf:1996iy}) lead to different results. Such differences can be interpreted as different choices of counterterms involving local higher dimension (holomorphic) operators that appear in the OPE of the spin-three currents with each other and with the twist fields. We apply our physically motivated prescription to compute the R\'enyi entropies. Our procedure also applies to contact terms arising from correlators of holomorphic and anti-holomorphic operators and yields a vanishing result for the corresponding integrals.

We find that the order $\mu^2$ correction to the R\'enyi entropies are given as,
\be
S^{(n)}_2= \frac{5c\mu^2}{6\pi^2\,(n-1)}\,(f_1\,{\cal I}_1\, + \,f_2\,{\cal I}_2)\,,
\ee
with
\bea
&&{\cal I}_1\,=\, \tfrac{4\pi^4}{3\beta^2}\,\left(\tfrac{4\pi\Delta}{\beta}\,\coth\left(\tfrac{\pi\Delta}{\beta}\right)\,-\,1\right)\,+\,
\label{I1final}\\\nonumber
&&+\tfrac{4\pi^4}{\beta^2\sinh^{2}\left(\tfrac{\pi\Delta}{\beta}\right)}\,\left\{\left(1-\tfrac{\pi\Delta}{\beta}\coth\left(\tfrac{\pi\Delta}{\beta}\right)\right)^2\,-\,\left(\tfrac{\pi\Delta}{\beta}\right)^2\right\}\\\nonumber
&&{\cal I}_2\,=\,\tfrac{8\pi^4}{\beta^2}\,\left(5\,-\,\tfrac{4\pi\Delta}{\beta}\,\coth\left(\tfrac{\pi\Delta}{\beta}\right)\right)\,+\,
\label{I2final}\\\nonumber
&&+\tfrac{72\pi^4}{\beta^2\sinh^{2}\left(\tfrac{\pi\Delta}{\beta}\right)}\,\left\{\left(1-\tfrac{\pi\Delta}{\beta}\coth\left(\tfrac{\pi\Delta}{\beta}\right)\right)^2-\tfrac{1}{9}\left(\tfrac{\pi\Delta}{\beta}\right)^2\right\}\,.
\eea
In the limit $n\to 1$, and at high temperatures ${\Delta\beta^{-1}}\gg 1$ these two terms correctly reproduce the order $\mu^2$ correction to the thermal entropy 
\eqref{thermal}, as expected from the extensivity of high temperature EE. 

A striking confirmation of the universality of the order $\mu^2$ correction to higher spin EE argued above, is provided by the spin three black hole solution of 
\cite{gutkraus} using the holographic EE proposal of \cite{dbj}. The higher spin black holes in AdS$_3$ have been constructed in the ${\rm SL}(3,{\mathbb R})\times{\rm SL}(3,{\mathbb R})$ Chern-Simons formulation of higher spin gravity. The EE in the boundary CFT$_2$ with ${\cal W}_3$-symmetry is computed by a Wilson line in the ${\rm SL}(3)$ Chern-Simons theory connecting the end-points of the interval in the boundary CFT. Explicit computation of the order $\mu^2$ correction has shown precise agreement 
\cite{paper1} with the results for EE argued above. It would be extremely interesting to understand if this agreement also extends to R\'enyi entropies computed from first principles in the bulk gravity dual. An important point to note is that the universality in the single-interval RE/EE we have shown in this paper will not extend to higher orders in $\mu$. This is because the entanglement entropy must reduce to the thermal entropy for large intervals and it is known that the higher order corrections to thermal entropy depend non-trivially on $\lambda$ for theories with ${\cal W}_\infty[\lambda]$ symmetry (see e.g. \cite{fromCFT}).

In this paper we have shown that the four-point correlator involving branch-point twist fields and spin-three currents is universal. 
The results presented here can be reproduced by applying the uniformization map to two-point correlators on the plane and summing over Riemann sheets appropriately. This line of reasoning will be presented elsewhere. It would be interesting to find examples other than free boson and free fermion CFTs where the explicit computation of this correlator is possible.  The methods employed in this paper can be used more generally to study deformations of CFTs by holomorphic operators. For example, it is possible to compute leading corrections to RE/EE with chemical potentials for charges of generic spin in CFTs with ${\cal W}$-symmetry, and the results can be compared with holographic methods applied to corresponding higher spin black holes.

Finally, we remark on the relation of our work to previous studies of entanglement and R\'enyi entropies in perturbed CFTs, e.g. in \cite{Cardy:2010zs}. In particular, 
\cite{Cardy:2010zs} dealt with perturbing operators which were primary fields but with weights $(h, \bar h)$ where $\bar h \neq 0$. For the situation considered in this paper the perturbation is holomorphic (plus antiholomorphic), the conformal perturbation theory is considerably easier to handle and exact results could be obtained .  Another important difference is that the perturbing operator considered in this paper is a perturbation of the basic Lagrangian of the theory. Therefore when evaluating the entanglement entropy using the replica trick, the perturbing operator is present in each copy of the CFT which implies these operators are singlets under the cyclic permutation of the copies. This is unlike the situation considered in \cite{Cardy:2010zs}  where the operators were not singlets under cyclic permutation of the replica copies.

The authors would like to thank Tim Hollowood for enjoyable discussions. The work of JRD is partially supported by the Ramanujan fellowship DST-SR/S2/RJN-59/2009. MF and SPK are supported by STFC grant
ST/J00040X/1.

\end{document}